\documentstyle[prl,aps,multicol,epsfig]{revtex}

\begin{document}

\title{\bf Entropic analysis of the role of words in literary texts}

\author{Marcelo A. Montemurro}
\address{
Facultad de Matem\'atica, Astronom\'{\i}a y F\'{\i}sica,
Universidad Nacional de C\'ordoba \\
Ciudad Universitaria, 5000 C\'ordoba, Argentina}

\author{Dami\'an H. Zanette}
\address{
Consejo Nacional de Investigaciones Cient\'{\i}ficas y
T\'ecnicas \\ Centro At\'omico Bariloche and Instituto Balseiro,
8400 Bariloche, R\'{\i}o Negro, Argentina}

\maketitle

\begin{abstract}
Beyond the local constraints imposed by grammar, words
concatenated in long sequences carrying a complex message show
statistical regularities that may reflect their linguistic role
in the message. In this paper, we perform a systematic statistical
analysis of the use of words in literary English corpora.  We
show that there is  a quantitative  relation between the role of
content words in literary  English and the Shannon information
entropy defined over an appropriate probability distribution.
Without assuming any previous knowledge about the syntactic
structure of language, we are able to cluster certain groups of
words according to their specific role in the text.
\end{abstract}

\begin{multicols}{2}


Language  is  probably the  most  complex  function of our brain.
Its evolutionary  success  has  been  attributed  to the high
degree of combinatorial power derived  from its fundamental
syntactic structure \cite{nowak}. Syntactic rules act locally at
the sentence level and do not necessarily  account for  higher
levels  of  organisation  in  large sequences of words conveying
a  coherent message \cite{akmajian}. In this respect,  the
situation is  similar to  that found  in other natural sequences
with non-trivial information content, such as the genetic code,
where  more than one structural level may  be discerned
\cite{daune,smith}. In the case of human language, complex
hierarchies have been revealed  at levels ranging from word form
to sentence structure \cite{akmajian,pinker1,pinker2}. Moreover,
it has  been argued that long samples of continuous written or
spoken language also possess a hierarchical macrostructure at
levels beyond the sentence \cite{dijk}. In a coarse grained
splitting of this complex organisation we could distinguish three
basic structural levels in the analysis of long language records.
The first  one corresponds to the absolute quantitative
occurrence of words, that is, which words are used and how many
times  each.  The second  level of organisation refers  to  the
particular  ways in which words can be linked into sentences
according to the syntactic rules of  language. Finally, at the
highest level, grammatical  sentences are combined in order to
thread a meaningful messages as part of a communications process.
This assemblage of sentences into more complex  structures  is
not strictly  framed by a set of precise prescriptions  and is
more related to the particular nature of the message conveyed by
the sequence. In this paper we shall focus on  the statistical
manifestations  of this  high level of organisation in
language.   By means of an entropic measure of word distribution
in literary corpora, we show that the statistical realisation of
words within a complex communicative structure reflects
systematic patterns which can be used  to cluster words according
to their specific linguistic role.


Zipf's  analysis \cite{Zipf}  represents the   crudest
statistical approach   by which some quantitative information
about the use of words in a corpus of written language can be
obtained. Basically, it consists of counting the number of
occurrences of each different word in  the  corpus,  and then
producing  a list of these words sorted according to decreasing
frequency. The rank-frequency  distribution thus obtained
presents robust quantitative regularities that have been tested
over a vast variety of natural languages. However, the
frequency-ordered list alone bears little information  on the
particular role of words in the lexicon, as can be realised by
noting  that  after shuffling  the corpus  the rank-frequency
distribution remains intact. Naturally, the first ranks in the
list belong to  the commonest  words in  the language  style of
the source text, e.g.  function  words  and some  pronouns in
literary English. After them, words related to the particular
contents of the text start to appear. An illustration is given
in  Table \ref{Hamlet}, where we show some portions of the first
ranks in Zipf's classification of the words from William
Shakespeare's Hamlet.

It  is  therefore clear that in order to extract information about
the specific role of words by statistical analysis, we must be
able to gauge not only how often a word is used but also where it
is used  in  the  text.  A  statistical measure that fulfills the
aforementioned  requirement  can  be  constructed  from a suitable
adaptation  of the Shannon information entropy \cite{shannon}. Let
us  think  of  a given text corpus as made up of the concatenation
of  $P$  individual  parts. The kind of partitions we are going to
consider  here  are those that arise naturally at different scales
as  a  consequence  of  the  global structure of literary corpora.
Examples  of  these  natural divisions are the individual books of
an  author's whole production, and the collection of chapters in a
single  book. Calling $N_i$ the total number of words in part $i$,
and  $n_i$ the number of occurrences of a given word in that part,
the  ratio  $f_i=n_i/N_i$ gives the frequency of appearance of the
word in question in part $i$. For each word, it is possible to
define a probability measure $p_i$ over the partition as
\begin{equation}
p_i =\frac{f_i }{\sum_{j=1}^P f_j },
\end{equation}
The quantity $p_i$ stands for the probability of finding the word
in  part $i$, given  that  it is present  in the corpus. The
Shannon information entropy associated with the discrete
probability distribution $p_i$ reads
\begin{equation}
\label{entropy}
S=-\frac{1}{\ln P}\sum_{i=1}^P p_i\ln p_i .
\end{equation}
Generally,  the  value  of  $S$  is  different  for  each word. As
discussed   below,   the  entropy  of  a  given  word  provides  a
characterization  of  its  distribution  over  the different
partitions. Note that,  independently  of  the  specific  values
of $p_i$, we have $0\leq S \leq 1$.

To  gain  insight  on  the  kind  of  measure  represented by $S$,
two  limiting  cases  are  worth  mentioning.  If  a given word is
uniformly  distributed over the $P$ parts,   $p_i=1/P$ for all
$i$  and  equation  (\ref{entropy}) yields $S=1$. Conversely, if a
word  appears  in  part  $j$  only,  we  have $p_j=1$ and $p_i =0$
for   $i\neq   j$,  so  that  $S  =0$.  These  examples  represent
extreme   real   cases   in   the  distribution  of  words.  In  a
first  approximation  one  expects  that  certain words are evenly
used  throughout  the  text regardless of the specific contents of
the   different  parts.  Possible  candidates  are  given  by
function words, such as articles and prepositions,  whose  use is
only weakly affected by the specific  character  of  the different
parts in  a homogeneous corpus.  Other  words, associated with
more particular aspects of each part may  fluctuate considerably
in their use, thus   having  lower values  of the entropy.  We
show  in  the following  that  in just  a few statistical
quantities such as frequency  and entropy there is relevant
information about the role of certain word classes.

\begin{figure}[b]
\centerline{\psfig{file=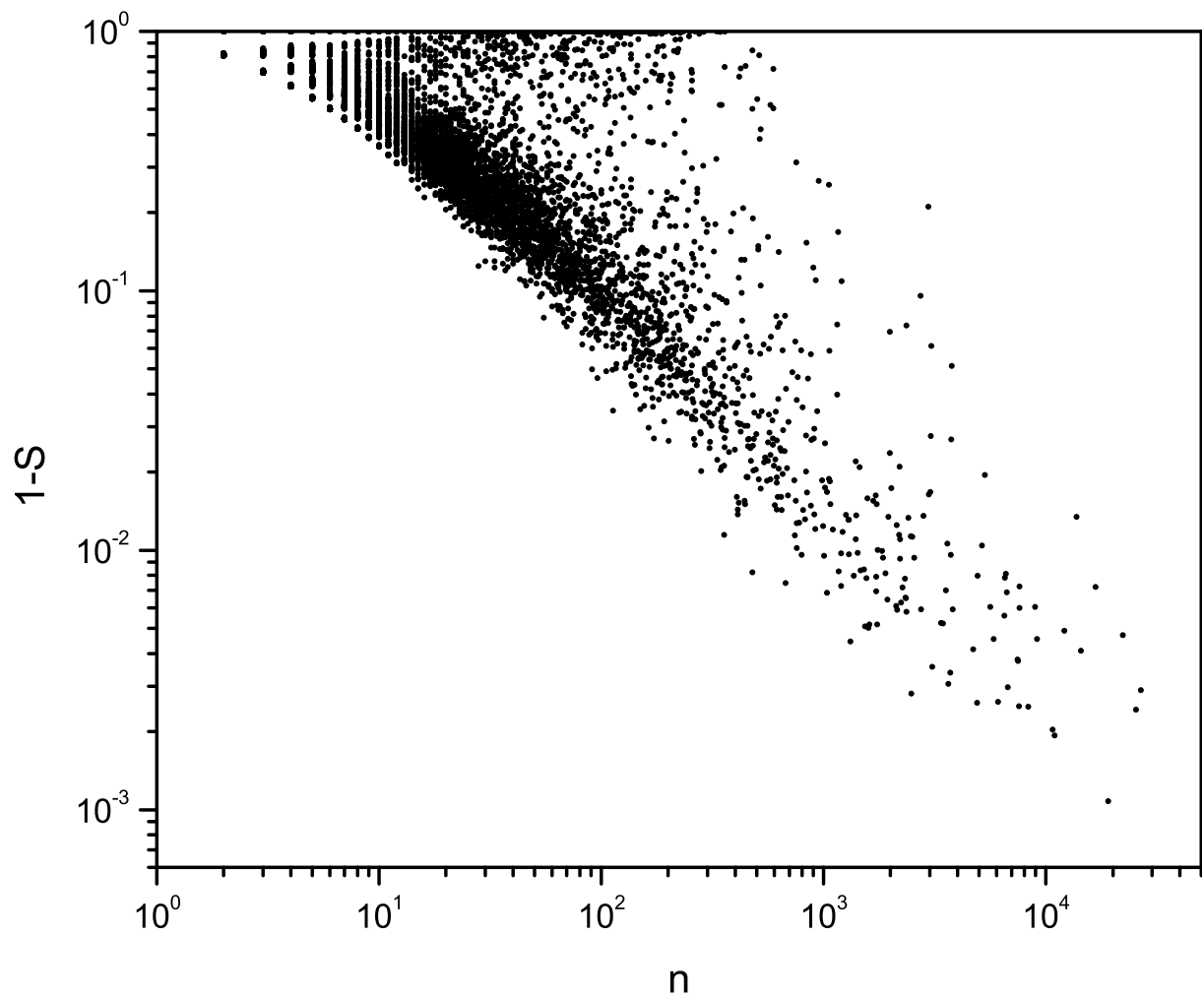,width=.9\columnwidth}}
\caption{Plot of $1-S$ versus the number of occurrences $n$  for
each word of a corpus made up of 36 plays by William Shakespeare.
The total number of words is 885,535 and the number of different
words is 23,150.}
\label{f1}\end{figure}

Figure    \ref{f1}   shows   $1-S$,   with   $S$   calculated   as
in  equation  (\ref{entropy}),  versus  the  number of occurrences
$n$,   for  each  different  word  in  a  corpus  made  up  of
$36$ plays  by  William  Shakespeare.  The  total  number  of
words in the    set    of    plays    adds    up    to $885,535$
with   a vocabulary   of   $23,150$   different  words. In  this
case,  the natural   division  that  we  are considering  is
given  by  the individual  plays.  The structure  of  the  graph
calls  for two different   levels of   analysis.   First,   its
most  evident feature,   that is  the  tendency  of  the entropy
to  increase with   $n$, represents   a  general  trend of  the
data  which should   be explained  as  a  consequence of  basic
statistical facts. In   qualitative   terms,  it implies  that
on  average the more  frequent  a  word  is  the more  uniformly
is it used. Second, a  somewhat  deeper  quest may  be  required
in order to reveal whether  the  individual deviations  from
this  general trend are  related  to  the particular  usage
nuances of words, as imposed  by  their specific  role  in  the
text. Whereas most methods of  word clustering  according  to
predefined  classes heavily  rely  on a  certain  amount  of
pre-processing, such as tagging  words as  members  of particular
grammatical categories \cite{pereira},  we  shall address  this
point without any {\em a priori}  linguistic knowledge,  save
the  mere identification of words as the minimal structural units
of language.

\begin{figure}[b]
\centerline{\psfig{file=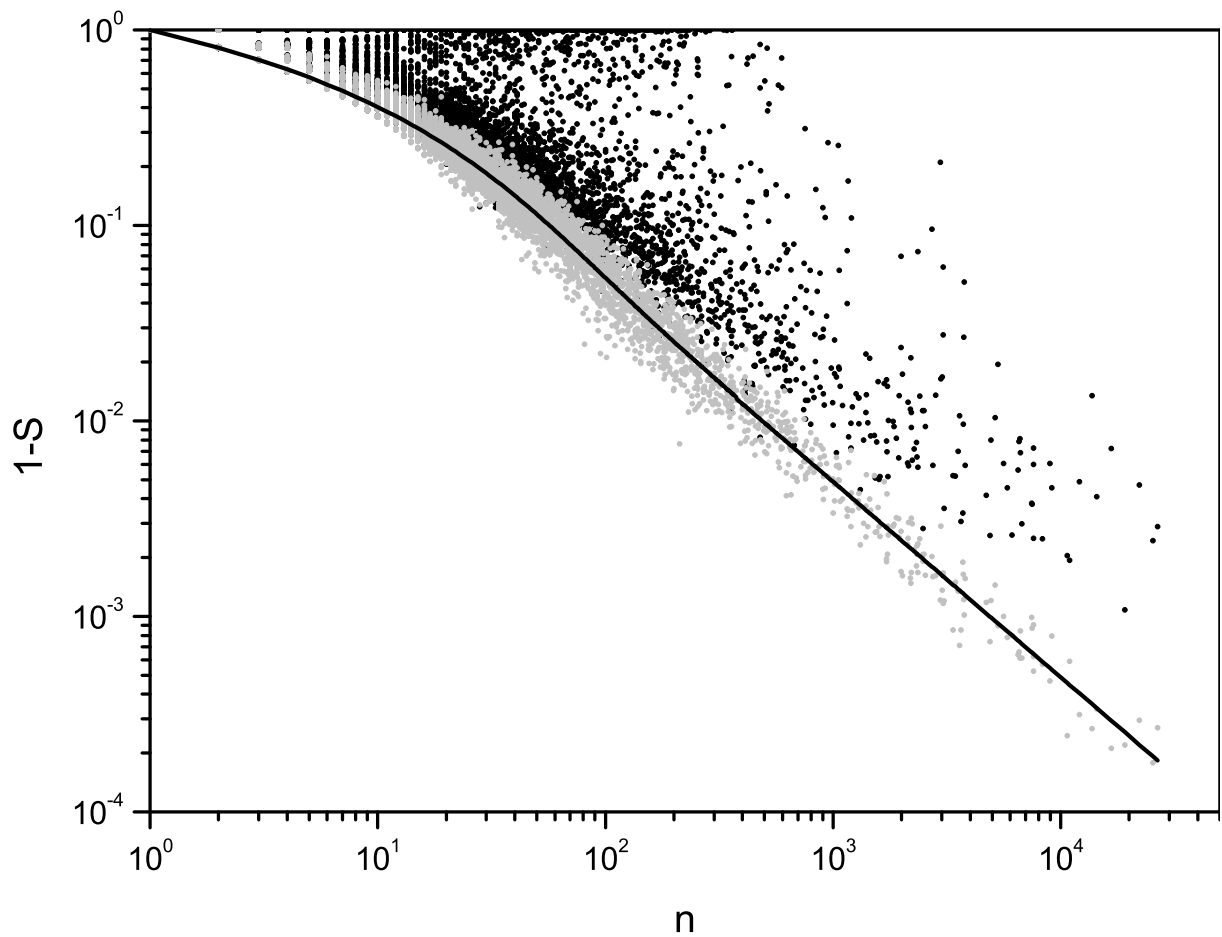,width=.9\columnwidth}}
\caption{Comparison between the  data shown in Figure \ref{f1}
(black dots) and a randomised version of the Shakespeare corpus
(grey dots). The curve stands for the analytical approximation for
the random corpus, equation (\ref{meanS}).}
\label{f2}\end{figure}

In  order  to  clarify  to  which  extent the features observed in
figure   \ref{f1}   reflect   basic   statistical   properties  of
the  distribution  of  words  over  the  different  parts  of  the
corpus,   we   performed   a  simple  numerical  experiment  which
consists   in   generating  a  {\em  random  version}  of  the
$36$ Shakespeare's  plays.  This  was  done  in  the  following
steps. First,  we  considered  a  list  of  all  the  words used
in the plays,   each  appearing  exactly  the  number  of times
it  was used  in  the  real  corpus.  Second,  we shuffled  the
list thus completely  destroying  the  natural order  of  words.
Third, we took  the  words  one  by  one from  the  list  and
{\em wrote} a random    version    of each    play containing
the   same number   of   words as   its   real counterpart. In
figure \ref{f2},  we compare  the randomised  version  with the
data of figure \ref{f1}.  It  is evident that, on one hand, the
tendency of the  entropy  to grow with $n$ is preserved. On the
other, the large fluctuations in the value of $S$, as well as the
presence of relatively infrequent  words  with  very low entropy,
are totally erased in  the  randomised  version.  On average,
words have  higher   entropies in the random realisation   than in
the  actual corpus. Indeed,  this is what  one  would expect for
certain word classes such as proper nouns  and,  in general, for
content words that allude to objects, situations or  actions
related  to specific parts of the corpus. All  the
inhomogeneities  that characterise the use of such words
disappear in   the   random version,  and consequently  render
higher values of the entropy.

Besides  its  value as a comparative benchmark, the random version
of  the  corpus has the appeal of being analytically tractable, at
least  in  a  slightly  modified  form, as follows. Let us suppose
that  we  have a corpus of $N$ words consisting of $P$ parts, with
$N_i$  words  in  part $i$ ($i=1,\dots,P$). The probability that a
word  appears  $n_1$  times  in part $1$, $n_2$ times in part $2$,
and so on, is
\begin{equation}    \label{prob}
p(n_1,n_2,\dots ,n_P) = n! \prod_{j=1}^P \frac{1}{n_j!} \left(
\frac{N_j}{N} \right)^{n_j},
\end{equation}
with $n=\sum_j  n_j$. In  the special  case where  all the  parts
have exactly the  same number  of words,  i.e. $N_i=N/P$  for
all  $i$, the average value of the entropy  resulting from the
probability given  by equation (\ref{prob}) can be written in
terms of $n$ only, as
\begin{equation}
\label{meanS}
\langle S(n)\rangle=- \frac{P}{\ln P}\sum_{m=0}^n \frac{m}{n} \ln
\left( \frac{m}{n} \right) {{n} \choose{m}}
\frac{1}{P^m}\left(1-\frac{1}{P}\right)^{n-m} .
\end{equation}
For highly frequent words, $n\gg 1$, equation (\ref{meanS})
assumes a particularly simple form, namely
\begin{equation} \label{asymp}
\langle S(n) \rangle \approx 1-\frac{P-1}{2 n \ln P}  .
\end{equation}
The curve in figure \ref{f2} stands for the function $1-\langle
S(n) \rangle$, with $\langle S(n) \rangle$ given by equation
(\ref{meanS}) with $P=36$. First, we note that despite  the  fact
that $\langle  S(n) \rangle$ was calculated assuming  that  all
the parts  have the same number of words, its agreement with the
random realisation for all the frequency range  is very good.
Moreover, it can be seen that after a short transient  in the
region  of low  $n$, $1-\langle S(n) \rangle$  soon develops  the
asymptotic form given by equation (\ref{asymp}) --a straight line
of slope $-1$ in this log-log plot.

\begin{figure}[b]
\centerline{\psfig{file=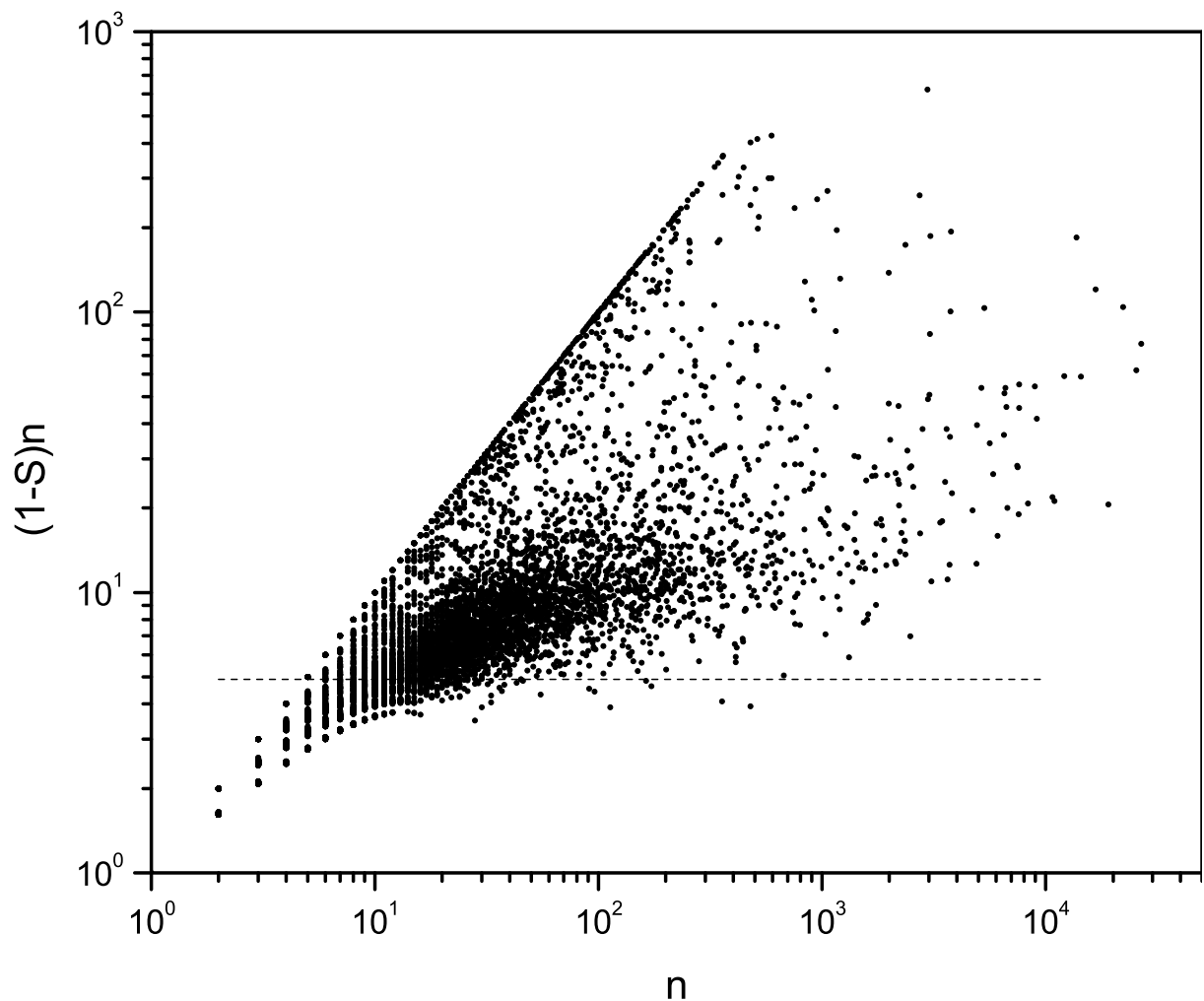,width=.9\columnwidth}}
\caption{Plot of $(1-S)n$ vs. $n$ for all the different words in
the  Shakespeare corpus. The horizontal line shows the expected
value of $(1-S)n$ for frequent, uniformly distributed words, as
given by equation (\ref{filter}).}
\label{f3}\end{figure}

We have so  far explained the  general trend in  the behaviour
of  the entropy  with   simple  statistical   considerations
pertaining   the distribution of words over the different parts
of the corpus.  In  the second part  of our  analysis we  shall
address  the more  interesting question of how  the information
contained  in the entropy  may reveal natural groupings of
English words according to their particular  role in the text.
In  order to accomplish this  task it proves useful  to define
adequate {\em coordinates}  whereby words can be  associated to
points  in  a  suitable  space.    As  a  consequence  of that,
the classification  of  words  according  to  their role should
emerge naturally from the preference of certain words to occupy
more or  less definite regions of that space. As one of these
coordinates we take  a quantity reflecting the degree of use  of
a word in the text,  namely, the number of occurrences $n$. The
second coordinate is introduced  to measure the deviation  of
the  entropy  of each word from the value predicted by the
random-corpus model,  as follows.  We have  seen that equation
(\ref{asymp}) accounts for the expected statistical  decrease in
the  value of  the entropy  as a  word becomes  less frequent.
This effect can be separated  from the behaviour of  the words in
the  real texts, in order to reveal genuine information on the
linguistic  usage of words. We rewrite equation  (\ref{asymp}) as
\begin{equation}
\label{filter}
(1-\langle S \rangle ) n \approx  \frac{P-1}{2\ln P} ,
\end{equation}
where the right-hand side is independent of $n$. Therefore, in a
graph of $(1-S)n$  versus $n$,  the words  whose actual
distribution agrees with  the  random-corpus  model  should
approximately  fall  along  a horizontal line. All appreciable
departures from this line  should be expected to  bear some
relation to  the non-random  character of  the usage  of words,
and  therefore  may   reflect  actual   linguistic information.
By means  of relation  (\ref{filter}) we  are therefore able to
filter out all the trivial part of the statistical  behaviour.
Figure  \ref{f3} shows actual data for the Shakespeare corpus in
a plot  of $(1-S)n$ versus $n$. The horizontal  line stands for
the  value given in the right-hand side of equation
(\ref{filter}) for $P=36$.

In order to reveal whether the words show  some sort of
systematic distribution over the  plane according  to their
linguistic role, we proceeded to classify by hand the first
$2,000$ words into different sets.  The classification stops there
due to the fact that words close to that  rank occur in the whole
corpus a number of times similar to the number  of parts in the
corpus division, $n\approx P$, thus representing  a limit beyond
which statistical fluctuations start to dominate. The six
categories we set out in groups were the following: (a) proper
nouns, (b) pronouns, (c) nouns referring to humans, such as {\it
soldier} and {\it brother}, (d) nouns referring to nobility
status --such as {\it King} and {\it Duke}, which have a relevant
place in Shakespeare's plays--  (e) common nouns (not referring
to humans or to nobility status) and adjectives,   and finally
(f) verbs and adverbs. In case of ambiguity about the inclusion
of a word into a certain class we simply left it out and did not
classify it, hence the total number of classified words was
finally around $1,400$.

\begin{figure}[b]
\centerline{\psfig{file=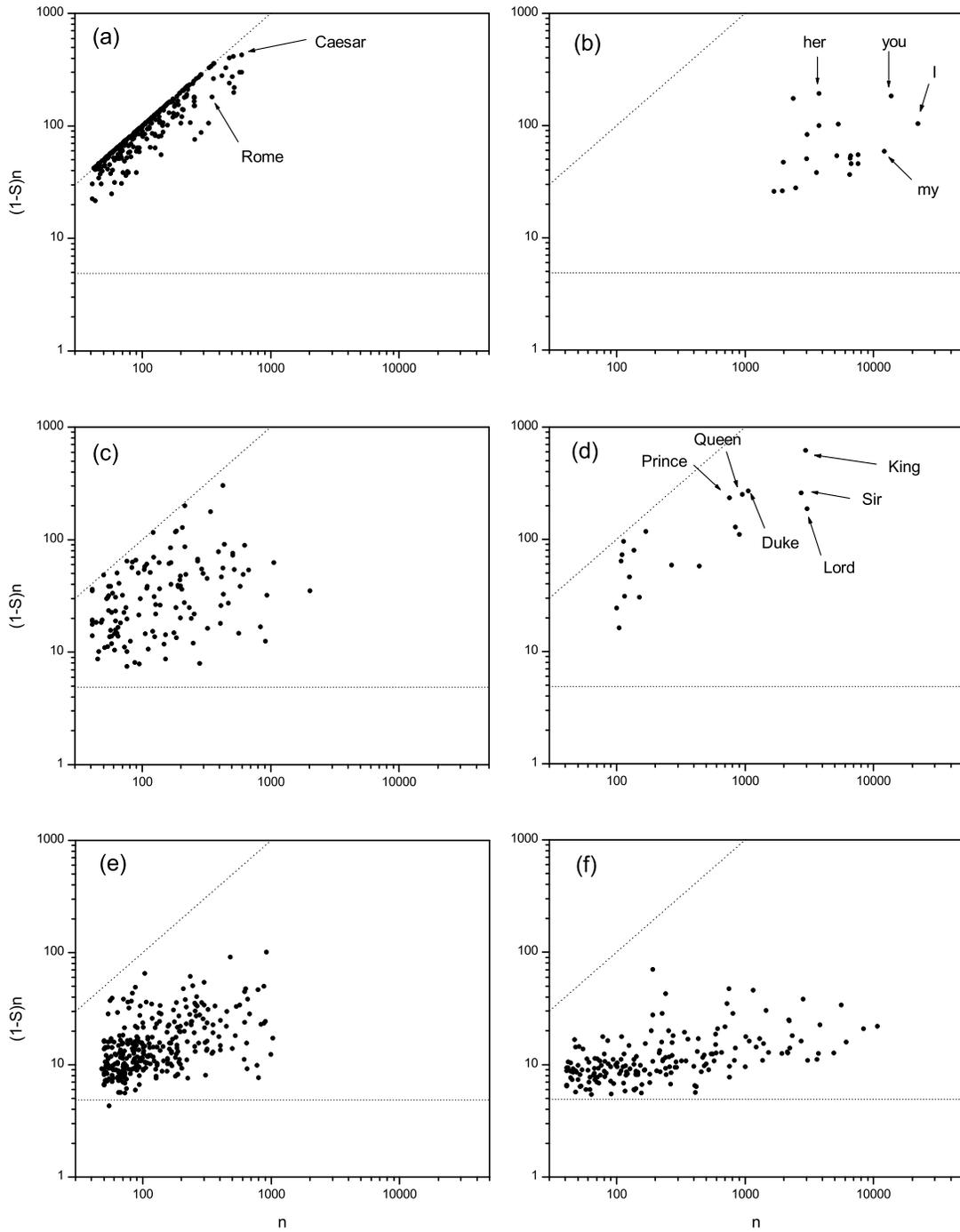,width=.8\columnwidth}}
\caption{Plot of $(1-S) n$ vs. $n$ for six relevant word
classes: (a) proper nouns, (b) pronouns, (c) nouns referring to
humans, (d) nouns referring to nobility status, (e) common nouns
(not referring to humans or to nobility status) and adjectives,
and (f) verbs and adverbs. In each plot, the horizontal dotted
line stands for the asymptotic value of $(1-S)n$ for the
random-corpus model, equation \ref{asymp}. Words close to this
line are homogeneously distributed over the corpus. The oblique
dotted line corresponds to $S = 0$. Proximity to this line
indicates extreme inhomogeneity in the distribution.}
\label{f4}\end{figure}

The results of this classification can be seen in Figure \ref{f4},
and in fact reveal a marked clustering of words over definite
regions of the two dimensional space spanned by $(1-S)n$ and $n$.
The sharpest distribution, shown in Figure \ref{f4}a, corresponds
to proper nouns. These words occupy a dense and elongated region
which is limited from above by the straight line representing the
identity function $(1-S)n=n$. Naturally, proper nouns are
expected to define a class of words strongly related to particular
parts of the corpus. In consequence, their entropies tend to be
very low on average, if not strictly zero as in the case of many
proper nouns appearing in just one of the Shakespeare's plays.
Thereby, in a graph of $(1-S)n$ versus $n$, words having values
of the entropy close to zero have $(1-S)n \approx n$ and fall
close to the identity function.

\begin{figure}
\centerline{\psfig{file=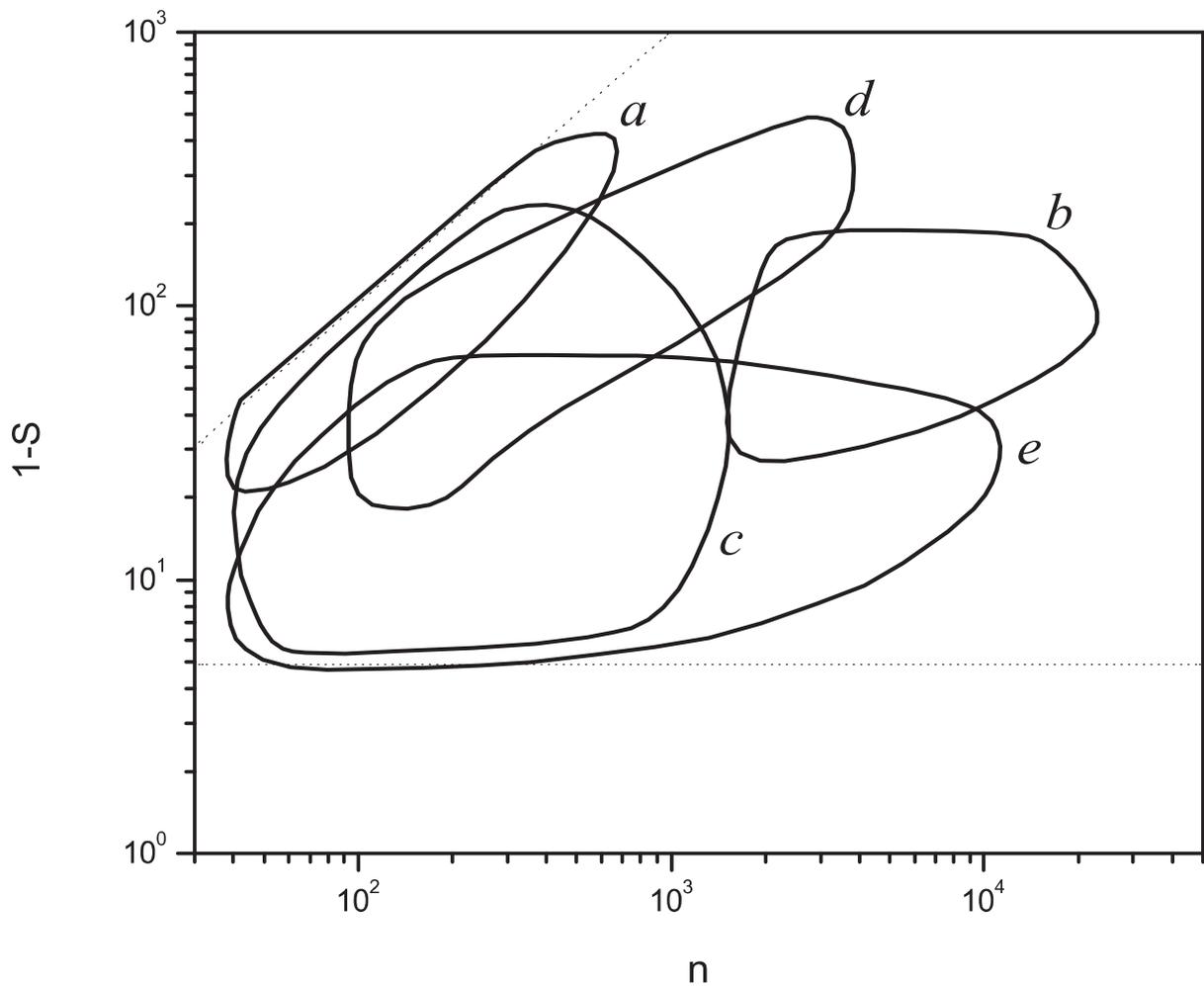,width=.9\columnwidth}}
\caption{Schematic combined representation of the zones occupied by the
word classes of figure \ref{f4}: proper nouns ({\it a}); pronouns
({\it b}); common nouns and adjectives, including those referring
to humans but not to nobility status ({\it c}); nouns referring to
nobility status ({\it d}); verbs and adverbs ({\it e}).}
\label{f5}\end{figure}

The distribution of other word classes is less obvious. Verbs and
adverbs (Fig. \ref{f4}f) are closest to the random distribution,
covering a wide range of ranks. On average, common nouns and
adjectives (Fig. \ref{f4}e) are farther from the random
distribution and, at the same time, are less frequent. Nouns
referring to humans (Fig. \ref{f4}c) cover approximately the same
frequencies as common nouns, but their distribution is typically
more heterogeneous. The entropy of some words in this class is,
in fact, quite close to zero. The three most frequent nouns in
the Shakespeare corpus are {\it Lord}, {\it King}, and {\it Sir}.
All of them belong to the class of nouns referring to nobility
status (Fig. \ref{f4}d), which spans a large interval of
frequencies and has systematically low entropies. The specificity
of nobility titles with respect to the different parts of the
corpus can be explained with essentially the same arguments as
for proper nouns. Considerably more surprising is the case of
pronouns (Fig. \ref{f4}b) which, as expected, are highly
frequent, but whose entropies reveal a markedly nonuniform
distribution over the corpus. The origin of this heterogeneity in
the distribution of pronouns is not at all clear, and deserves
further investigation. In Fig. \ref{f5} we have drawn together
the zones occupied by all the classes to make more clear their
relative differences in frequency and homogeneity.


We have performed the same statistical analysis over other
literary corpora, obtaining totally consistent results. The same
organisation of words was observed in the works of Charles
Dickens and Robert Louis Stevenson. In particular, nouns and
adjectives tend to be more heterogeneously distributed than verbs
and adverbs. Nouns referring to humans have systematically lower
entropies. Pronouns, in turn, exhibit an unexpectedly
heterogeneous distribution for their high frequencies.

In summary, in this work we have concentrated on the statistical
analysis of language at a high level of its structural hierarchy,
beyond the local rules defined by sentence grammar. We started
off by introducing an adequate measure of the entropy of words in
a text corpus made up of a number of individual parts. With
respect to Zipf's analysis, which focuses on the frequency
distribution of words, the study of entropy provides a second
degree of freedom that resolves the statistical behaviour of
words in connection with their linguistic role. By means of our
random-corpus model we were able to extract the nontrivial part of
the distribution of words. This procedure reveals statistical
regularities in the distribution, that can be used to cluster
words according to their role in the corpus without assuming any
{\em a priori} linguistic knowledge. Ultimately, such
regularities should stand as a manifestation of long-range
linguistic structures inherent to the communication process. We
believe that a thorough explanation of the origin of these global
structures in language may eventually contribute to the
understanding of the psycolinguistic basis for the modelling of
reality by the brain.


Critical reading of the text by Susanna Manrubia is gratefully
acknowledged.

\begin{table}
\begin{tabular}{l|r|r}
word & rank & number of occurrences\\
\hline
the &  1  & 1087\\
and & 2 & 968\\
to & 3 & 760\\
of & 4 & 669\\
I &  5 & 633\\
a &  6 & 567\\
you &   7 & 558\\
$\cdots$ & $\cdots$ & $\cdots$\\
Lord  & 25  &  225\\
he &  26 & 224\\
be   & 27 & 223\\
what  & 28 & 219\\
King & 29  & 201\\
him &  30  &  197\\
$\cdots$ & $\cdots$ & $\cdots$\\
Queen & 42 & 120\\
our &  43 & 120\\
if & 44 &  117\\
or &  45 &  115\\
shall  & 46 & 114\\
Hamlet &  47 & 112\\
$\cdots$ & $\cdots$ & $\cdots$ \\
\end{tabular}
\caption{Rank classification of words from
Shakespeare's Hamlet.} \label{Hamlet}
\end{table}

\end{multicols}

\end{document}